\documentclass[a4paper,11pt]{article}
\usepackage{pos}
\usepackage{doi}
\usepackage{sidecap}
\usepackage{hyperref}
\hypersetup{
 colorlinks=true,
 linkcolor=blue,
 filecolor=magenta, 
 urlcolor=cyan,
 pdftitle={Higgs in The Cosmos},
 pdfpagemode=FullScreen,
 }
\usepackage{amsmath}
\usepackage{slashed}
\usepackage{graphicx}
\usepackage{bm}
\usepackage{float}
\usepackage{subeqnarray}
\usepackage{eqnarray,amsmath}
\newcommand{\be}{\begin{equation}}
\newcommand{\beq}{\begin{equation}}
\newcommand{\ee}{\end{equation}}
\newcommand{\eeq}{\end{equation}}
\newcommand{\ba}{\begin{eqnarray}}
\newcommand{\ea}{\end{eqnarray}}
\newcommand{\ban}{\begin{eqnarray*}}
\newcommand{\ean}{\end{eqnarray*}}

\newcommand{\req}[1]{Eq.\,({\ref{#1}})}
\newcommand{\rf}[1]{Fig.\,{\ref{#1}}}
\newcommand{\rsec}[1]{Section\,{\ref{#1}}}
\newcommand{\orcJ}{0000-0001-8217-1484}
\newcommand{\orcC}{0000-0001-5038-8427}
\newcommand{\orcidicon}{\includegraphics[width=0.32cm]{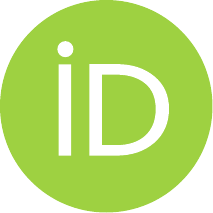}}
\newcommand{\orc}[1]{\href{https://orcid.org/#1}{\orcidicon}}

\title{Higgs in The Cosmos}
\ShortTitle{Higgs in The Cosmos}

\author*[a\orc{\orcJ}]{Johann Rafelski}
\author[a\orc{\orcC}]{Cheng Tao Yang}

\affiliation[a]{Department of Physics, 
 The University of Arizona, 
Tucson, Arizona 85721, USA}

\emailAdd{johannr@arizona.edu}
\emailAdd{chengtaoyang@arizona.edu}

\abstract{We explore the Higgs particle in the cosmic quark-gluon plasma (QGP) below the electroweak phase transition temperature $T_\mathrm{EW}\simeq 125\mathrm{\,GeV}$. We show that Higgs is neither in abundance (chemical) nor in momentum distribution equilibrium in certain stages of the Universe evolution. Nonequilibrium originates in: For chemical nonequilibrium in the always present irreversible decays into virtual heavy gauge bosons, and; For $T<25$\,GeV in relatively rapid $2\leftrightarrow 1$ formation and decay processes yielding momentum distribution as created in these reactions. As heavy particles disappear, the minimal Higgs coupling to abundant low mass particles fails in $2\to2$ (two-particle) scattering processes to assure a kinetic distribution equilibrium. The expansion of the Universe is by more than 10 orders of magnitude slower compared to microscopic processes. All other particles in the Universe are in full thermal equilibrium, with exception of the late in QGP evolution of the bottom flavor near to hadronization condition.
}
\FullConference{\today, Presented at The XVIth Quark Confinement and the Hadron Spectrum Conference\\ Cairns, Australia August 2024 % conference data here 
}

\begin{document}
\maketitle
%%%%%%%%%%%%%%%%%%%%%%%%%%%%%%%%%%%%%%%%%%%%%%%%%
\section{Baryogenesis, Nonequilibrium, and Heavy Particles in the Primordial QGP}\label{ssec:Kinetic}
The three conditions necessary to permit baryogenesis in the primordial Universe were formulated in 1967/1988 by Andrei Sakharov~\cite{Sakharov:1967dj, Sakharov:1988vdp}\\[-0.8cm]
\begin{subequations}\label{Sakharov}
\begin{align}
 &\text{Absence of baryon number conservation -- an obvious requirement;}\\[-0.2cm]
 &\text{Violation of CP-invariance -- to tell our Universe is made of `matter';}\\[-0.2cm]
&\text{Non-stationary conditions in absence of local thermodynamic equilibrium -- see below.}\\[-0.8cm]\nonumber
\end{align}
\end{subequations}
This work presents progress in our thorough exploration of possible nonequilibrium conditions in the primordial cosmic quark-gluon plasma (QGP) for all heavy particles: top (anti)quark $t,\bar t$, Higgs $h$, $W,Z$ gauge bosons; and recall of the case of bottom quark $b,\bar b$. We seek to recognize non-stationary conditions needed to form the baryon number surrounding us today. 

After 40+ years of research baryogenesis is still not understood. A possible explanation is that the precise epoch responsible for the baryonic matter genesis has not been established. Prior focus of baryogenesis research has been on the relatively short time interval near to the electroweak phase transition $T_\mathrm{EW}\simeq125\,\mathrm{GeV}$~\cite{Kuzmin:1985mm,Kuzmin:1987wn,Arnold:1987mh,Kolb:1996jt,Riotto:1999yt,Nielsen:2001fy,Giudice:2003jh,Davidson:2008bu,Morrissey:2012db,Canetti:2012zc}. However, a phase transition is not required for baryogenesis if non-stationary conditions were otherwise present in the primordial QGP. For followers of kinetic Boltzmann theory it is evident that in thermal equilibrium, the net effect of any baryogenesis processes is canceled by the equality between the forward and back reactions. Any non-stationary environment assures that the growth in baryon number is irreversible in a cosmological evolution that preserves entropy. We show here using kinetic reaction rates which we obtain, that the Higgs particle, as long as it is present, can catalyze non-stationary conditions in the Universe. In prior work we have demonstrated that this is also the case for the bottom (anti)quark~\cite{Rafelski:2024fej,Yang:2020nne} near to QGP hadronization.

Thermal equilibrium requires both chemical equilibrium in which particle abundances are at a maximum (`black body' yield, when speaking about photons) and kinetic equilibrium in which energy has been shared and distributed to maximize entropy, given the number of particles. In this work we show that for the Higgs particle, the abundance equilibrium is nearly always out of reach for the Higgs. In addition, the kinetic momentum distribution falls out of equilibrium as temperature $T\le 25$\,GeV, when the abundance of Higgs in the Universe is still quite significant.

The primordial cosmic QGP at temperature $125\,\mathrm{GeV}>T>0.15\,\mathrm{GeV}$ contained all the fundamental building blocks of matter; hadrons were dissolved into their constituent free quarks $u,d,s,c,b$ and gluons $ g$ with quark-antiquark abundance asymmetry containing seeds of present day baryon content of the Universe. The color deconfined matter, the quark-gluon plasma (QGP), is explored in laboratory experiments. However, the primordial plasma components such as charged leptons, photons, and neutrinos are not part of the thermal laboratory micro-bang studies: The tiny drops of QGP created in laboratory have a lifespan of around $10^{-22}$\,s --$10^{-23}$\,s, preventing electroweak (EW) processes from equilibrating. Similarly, the laboratory experiments do not probe any of the heavy particles. Therefore, a very careful theoretical analysis of the dynamical evolution of these heavy particles in the primordial Universe is required in order to asses any related non-equilibrium situation that could arises. The Higgs is here of particular interest as its pattern of coupling to other particles is unusual. When temperature decreases and heavy particles fade from the primordial QGP inventory, Higgs scattering on the lighter particle thermal background can diminish significantly, due to minimal coupling.

One can wonder how the common assumption of total equilibrium in the primordial cosmic QGP after electroweak phase transition has come to be. The reasoning is based on the magnitude of the characteristic Universe expansion time \begin{align}
\tau_\mathrm{U}\equiv 1/H\,,\quad H\equiv \dot a(t)/a(t)\,.
\label{HFriedman}
\end{align} 
Here $a(t)$ is the expansion scale of the Universe entering the cosmological FRWL model where $H$ and the energy density $\rho_i$ are related
\begin{align} 
H^2=\frac{8\pi G}{3}\left(\rho_\gamma+\rho_{\mathrm{lepton}}+\rho_{\mathrm{quark}}+\rho_{g,{W^\pm},{Z^0}}\right)\,.
\label{H2Friedman}
\end{align} 
Here $G$ is the Newtonian constant of gravitation. The Einstein cosmological constant-style dark energy is irrelevant in this primordial epoch as is dark matter in any form compatible with the present day dynamic Universe: both components are visible today due to extreme visible energy dilution of the Universe in subsequent expansion, see Fig.1 in Ref.\,\cite{Rafelski:2024fej}. The primordial QGP present in the early Universe during a temperature range of $125\,\mathrm{GeV} > T > 0.15\,\mathrm{GeV}$, is in this interval in the range $ \tau_\mathrm{U}=10^{-9}$\,s and $\tau_\mathrm{U}=10^{-5}$\,s, respectively. When this is compared to microscopic reactions it is clear that in comparison the Universe expansion is by 10 orders of magnitude too slow. However we found two exceptional situations, and the case of Higgs is described in this work. In both cases, bottom quarks being the other, non-stationary nonequilibrium relies on competing microscopic processes, where one is at least in part irreversible.

For what follows it is important to keep distinct the meanings of: 
a) detailed balance, 
b) thermal non-equilibrium - both chemical (abundance) and kinetic (phase space distribution), and 
c) non-stationary condition. 
Detailed balance between forward-backward reactions can be maintained even when {\it e.g.\/} particle abundance (chemical) equilibrium is not achieved. This is a stationary non-equilibrium state. In 1988 Sakharov excluded this reversible situation. The stationary and non-stationary non-reversible conditions can be separated as follows: By first considering as negligible the time dependence of the dynamic Universe (temperature $T(t)\to Const.$) we obtain a solution for the nonequilibrium we call adiabatic. The non-stationary component is found considering solution in presence of the dynamical expansion of the Universe ( $T(t)\ne Const.$). 

Using this approach we obtained the non-stationary component of the bottom flavor non-equilibrium, see chapter 2.3 in Ref.\,\cite{Rafelski:2024fej}, Figure 17. As our present study demonstrates~\cite{Yang:2025zuc}, the Higgs particle is another candidate for non-stationary dynamics across a much longer timespan of the primordial QGP phase. However, due to its being in non-equilibrium in both, the momentum distribution, and particle abundance the non-stationary component will require a more thorough study. Here we keep apart the meaning of~\cite{Koch:1986ud}: a) kinetic (momentum distribution) equilibrium and, b) chemical (particle abundance) equilibrium. At $T\gg m$ (the mass of particles involved) both equilibrium conditions are indistinguishable as we demonstrated in a kinetic study~\cite{Birrell:2014uka}. However, when $T\simeq m$, the kinetic equilibrium is usually established much more quickly, while abundance yields are more difficult to establish: This is so since for particles with masses in excess, or at least similar to ambient temperatures the creation processes need to overcome mass thresholds. 

The idea that chemical equilibrium could not be achieved in presence of slow particle production processes was first proposed by Bir\'o~\cite{Biro:1981zi} in the context of relativistic heavy ion collisions. The time dependent approach to equilibrium of strangeness in QGP was made explicit in~\cite{Rafelski:1982pu}. For a review of the early developments and study of the approach to chemical abundance equilibrium see~\cite{Koch:1986ud}, where also in chapter 6.3 the time dependent strangness abundance fugacity $\gamma$ was proposed: Chemical non-equilibrium can be described by introducing the pair abundance fugacity parameter now in general called $\Upsilon$ in the Fermi/Bose ($\pm$)-distribution~\cite{Letessier:1993qa} 
\begin{align}
f_{F/B}(\Upsilon_i,p_i)=\frac{1}{\Upsilon^{-1}_i\exp{\left[E(p_i)/T\right]}\pm1}\,,\qquad E(p_i)=\sqrt{m_i^2+p_i^2}
\,.
\label{quantumDist}
\end{align}
As temperature decreases the Boltzmann limit of these quantum distributions emerges since the exponential in the denominator dominates. In this limit $m>T$ particle yields are proportional to $\Upsilon$. In~\rf{HiggsDensity_fig} we see the thermal equilibrium $\Upsilon=1$ abundances of heavy particles in primordial QGP obtained from~\req{quantumDist}, the reference value used, the baryon density in the Universe, which is quantified in the following~\rsec{ssec:entro}. 

%FIGURE%%%%%%%%%%%%%%%%%%%%%%%%%%%%%%%%%%%%%%%%%%%%%%
\begin{figure}%[ht]
\centerline{\includegraphics[width=0.85\columnwidth]{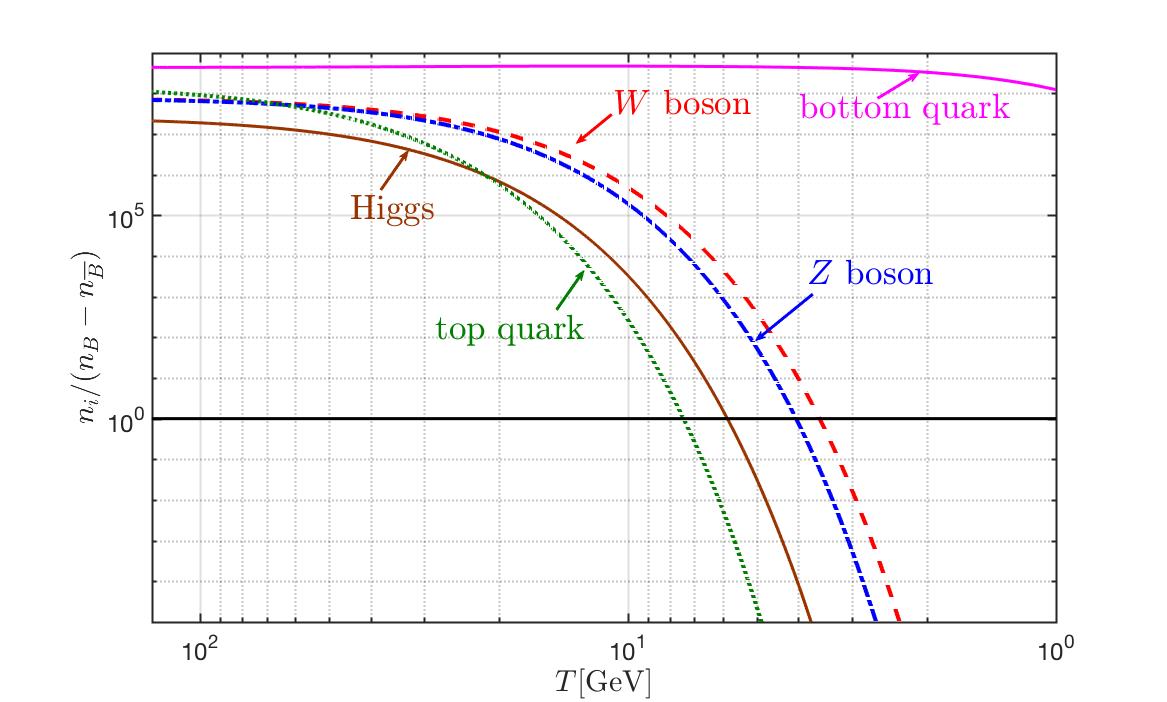}}
\caption{Thermal primordial QGP heavy particle abundance, presented as ratios with the baryon asymmetry, as a function of temperature $T$. For particle masses used see Ref.\,\cite{ParticleDataGroup:2024cfk}.}
\label{HiggsDensity_fig}
\end{figure}
%%%%%%%%%%%%%%%%%%%%%%%%%%%%%%%%%%%%%%%%%%%%%%

The Higgs minimal coupling has, for each particle of mass $m$, the value~\cite{ParticleDataGroup:2024cfk}
\begin{equation}
g_m=\frac{m\sqrt{2}}{v_0},\qquad v_0= 246 \,\mathrm{GeV}\,. 
\end{equation}
Here $v_o$ is the Higgs field vacuum field strength. As seen in~\rf{HiggsDensity_fig}, the bottom quark abundance persists: The bottom pair fusion process is scaling as $\propto g_m^2$, see~\rf{HiggsDiagram_fig} part (a), with $g_b=0.02$. Aside of the decay into bottom pair, decays into virtual heavy gauge mesons $W,Z$ with $g_Z=0.52$, $g_W=0.47$, must also be considered, see part (b)~\rf{HiggsDiagram_fig}. Higgs abundance equilibrium is broken since decay into two Gauge bosons of which one is virtual does not have a back reaction; the inverse $3\to 1$ process is of higher order in weak interaction and is suppressed by relevant coupling constants $g^2, g^{\prime\,2}$. This assures that Higgs abundance is always out of chemical equilibrium, which is different from the mechanism described in \cite{Vereshchagin:2017}.

The kinetic energy equilibration rate scales as $\propto g_m^4$, see part first two diagrams (${c_1}$) and (${c_2}$) in part (c),~\rf{HiggsDiagram_fig}, the third diagram (${c_3}$) relies on the triple Higgs coupling which in tree approximation is $\lambda_{hhh}^0=3m_h^2/v_0$, removing one power of $m_h$ for comparison with the Fermion exchange (${c_2}$) we see that that the effective coupling for Higgs exchange $g_{hhh}^\textrm{eff}\equiv \lambda_{hhh}^0/M_h=1.49 $, which is 1.5 times compared to top, since $g_t=0.99$. However, the scattering on light particles is small. As a consequence, the kinetic equilibration in the primordial QGP can be slower compared to production and decay rate after the more strongly coupled heavy $t, h, W, Z$ fade out below $T=25$\,GeV, see~\rf{HiggsDensity_fig}.

%FIGURE%%%%%%%%%%%%%%%%%%%%%%%%%%%%%%%%%%%%%%%%%%%%%%
\begin{figure}%[b]
\centerline{\includegraphics[width=0.78\columnwidth]{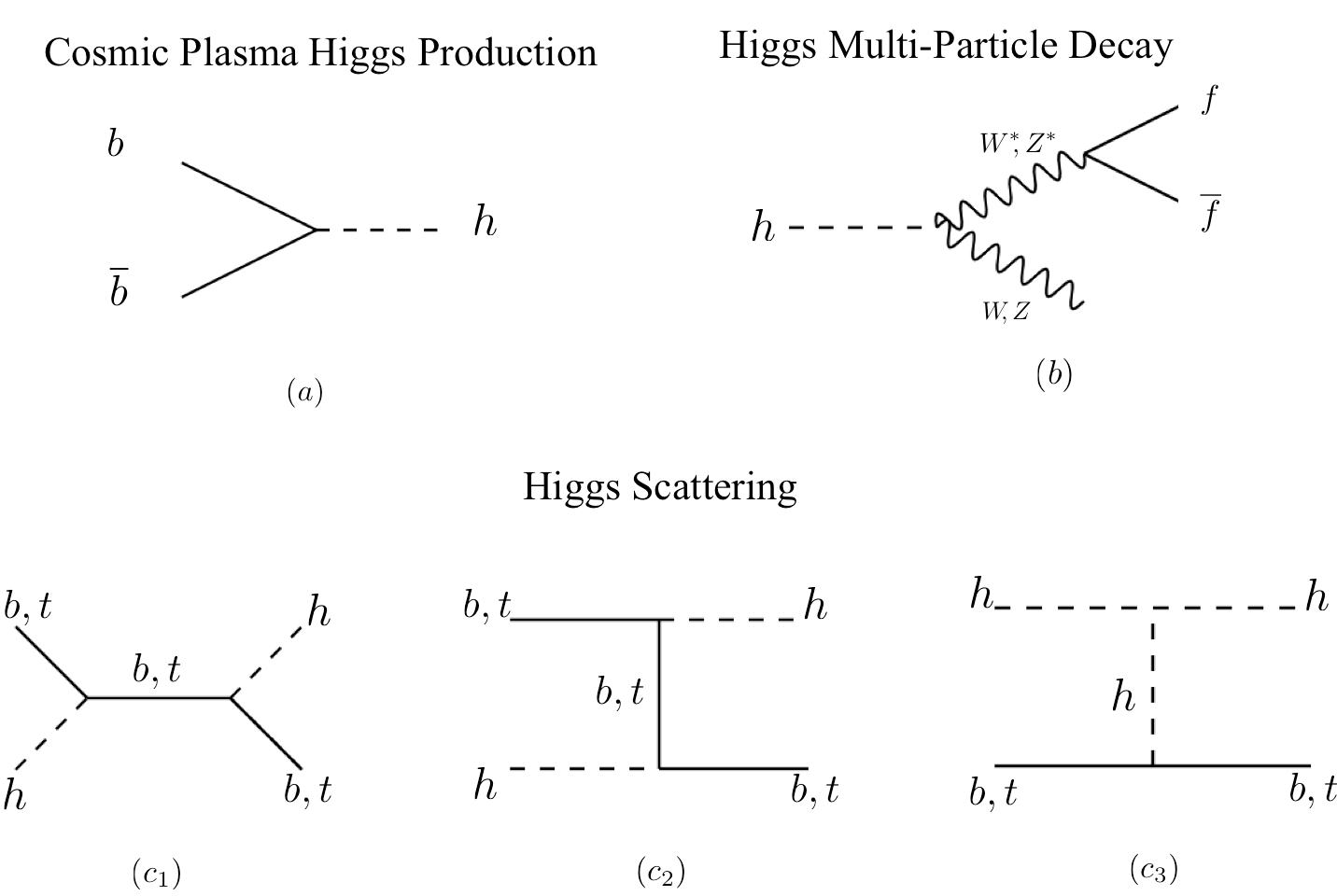}}
\caption{Lowest order Feynman diagrams in primordial QGP for:  $(a)$ dominant Higgs production $b\bar b\leftrightarrow h$; (b)  dominant Higgs  decay $h\to WW^\ast,ZZ^\ast$ and $(c)$ Higgs on $b,\bar b, t, \bar t$ quark scattering.\label{HiggsDiagram_fig}}
\end{figure}
%%%%%%%%%%%%%%%%%%%%%%%%%%%%%%%%%%%%%%%%%%%%%%%%%%%%%

In regard to the possible nonequilibrium of the yet heavier $ t, \bar t$ (anti)quarks, we note that the vertex coupling to the Higgs $g_{th}^2/4\pi\simeq 0.08$ is comparable to the coupling with the gluon, $\alpha_s\simeq 0.115$. Thus the number of vertices and the mass threshold determines which process dominates. Specifically, Higgs, gluon and quark-anti-quark fusion to form a top quark pair production processes are suppressed by the threshold $2m_t=350$\,GeV and the need to have two vertices. Therefore, even though EW coupling is 10 times weaker, in the primordial QGP chemical equilibrium is established by $ W+b\leftrightarrow t$ reaction, which is also the dominant top decay process $t\to W+b$, $\Gamma_t=1.4\pm0.2$\;GeV~\cite{ParticleDataGroup:2024cfk}. Strong interaction scattering on numerous quark components and gluons assure kinetic equilibrium. Once the gauge mesons disappear from the Universe inventory, any residual top abundance will disappear as well. Due to the coupling of gauge mesons to the high thermal abundance of all light particles this is expected at a very late stage $T\simeq m_W/40\simeq 2$\,GeV. At that point the residual heavy particle abundances are physically insignificant, see~\rf{HiggsDensity_fig}. As we noted above the other heavy particles, the gauge mesons $W,Z$ are sufficiently strongly coupled to the background high particle abundance so that we cannot expect a significant deviation from equilibrium. Thus aside the bottom only the Higgs creates in the primordial Universe an opportune context for non-stationary condition and `delayed' baryogenesis.

%%%%%%%%%%%%%%%%%%%%%%%%%%%%%%%%%%%%%%%%%%%%%%%%%%%%%%
\section{Baryon, Entropy, and Temperature Time Evolution in the Primordial Universe}\label{ssec:entro}
Since we use Einstein-Friedman cosmology, the entropy in the comoving volume $dV_c=a^3 dx^3$ is conserved. Following the era of baryogenesis, the baryon content in the comoving volume is also conserved. We can therefore relate the primordial cosmic QGP to present day using (subscript $t_0$ denotes the present day condition)
\begin{align}\label{BaryonEntropyRatio}
\left.\frac{n_B-n_{\overline{B}}}{s}\right|_{\mathrm{QGP}}= \left.\frac{n_B-n_{\overline{B}}}{ s_{\gamma,\nu}}\right|_{t_0}=\mathrm{Const.}=\left(\frac{n_B-n_{\overline{B}}}{n_\gamma}\right)\left(\frac{n_\gamma}{s_\gamma+s_\nu}\right)_{\!t_0}\!\!\!=(8.69\pm0.05)\times10^{-11}
\;.
\end{align}
To obtain the result shown we use:\\
\hspace*{0.8cm}i) The baryon-to-photon ratio 
\begin{equation}
\eta \equiv \left(\frac{n_B-n_{\overline{B}}}{n_\gamma}\right)=6.12 \times 10^{-10}\;.
\end{equation}
The latest reported 2024 value~\cite{ParticleDataGroup:2024cfk} is $\eta= 6.04(12) \times10^{-10}$; the value we have used is consistent.\\
\hspace*{0.8cm}ii) The entropy content today of free-streaming massless neutrinos $s_\nu$, and reheated photons $s_\gamma$~\cite{Birrell:2012gg}; electron-positron annihilation entropy is practically all flowing and contained in photons, hence 
\begin{align}
 \frac{s_\nu}{s_\gamma}=\frac{7}{8}\,\frac{g_\nu}{g_\gamma}\left(\frac{T_\nu}{T_\gamma}\right)^3\,,\qquad\frac{T_\nu}{T_\gamma}=\left(\frac{4}{11}\right)^{1/3}
 \,,
\end{align}
and the entropy-per-particle is
$s/n|_\mathrm{boson}\approx 3.60$,
$s/n|_\mathrm{fermion}\approx 4.20$ for massless bosons and fermions, respectively. The entropy density in QGP is defining the effective number of `entropy' degrees of freedom $g^s_\ast$ 
\begin{align}
 &s_{\mathrm{QGP}}=\frac{2\pi^2}{45}g^s_\ast T_\gamma^3\,,\qquad 
g^s_\ast=\!\!\sum_{i=\mathrm{bosons}}\!\!g_i\left({\frac{T_i}{T_\gamma}}\right)^3B\left(\frac{m_i}{T_i}\right)+\frac{7}{8}\sum_{i=\mathrm{fermions}}\!\!g_i\left({\frac{T_i}{T_\gamma}}\right)^3F\left(\frac{m_i}{T_i}\right).
\end{align}
The mass threshold functions for bosons $B(m_i/T)$ and fermions $F(m_i/T)$ are 
\begin{align}
&B\left(\frac{m_i}{T}\right)=\frac{15}{4\pi^4}\int^\infty_{m_i/T}\,dx\frac{\sqrt{x^2-\left({m_i}/{T}\right)^2}\left[4x^2-\left({m_i}/{T}\right)^2\right]}{\Upsilon^{-1}_ie^x-1}\,,\\
&F\left(\frac{m_i}{T}\right)=\frac{30}{7\pi^4}\int^\infty_{m_i/T}\,dx\frac{\sqrt{x^2-\left({m_i}/{T}\right)^2}\left[4x^2-\left({m_i}/{T}\right)^2\right]}{\Upsilon^{-1}_ie^x+1}\,.
\end{align}
Here $\Upsilon_i$ is the fugacity parameter for a given particle. 

When $T$ decreases below the mass of the particle ($T\ll m_i$) and becomes non-relativistic, the functions $B(m_i/T)$ and $F(m_i/T)$ go smoothly to zero, which implies that the contribution of the non-relativistic species to $g^s_\ast$ is negligible. The dominant factor $T^3$ cancels when computing the ratio of photons per entropy; therefore, the ratio of baryon number density to visible matter entropy density remains constant throughout the evolution of the Universe.

Since we are considering microscopic processes and their rates (in time), we need to determine the relation between time and temperature. To this end we differentiate the comoving entropy $S=\sigma V\equiv g^s_\ast T^3a^3=\mathrm{Const.}$ with respect to time to obtain 
\begin{align}
\left[\frac{\dot{T}}{g^s_\ast}\frac{dg^s_\ast}{dT}+3\frac{\dot{T}}{T}+3\frac{\dot{a}}{a}\right]g^s_\ast T^3a^3=0,\qquad \dot{T}=-\frac{HT}{1+\frac{T}{3g^s_\ast}\frac{d\,g^s_\ast}{dT}}=-HT\frac{1}{1+(1/3)d\ln g^s_\ast/d\ln T} \,.
\end{align}
This relation provides the quantitative solution for temperature as a function of time since $H$ is fixed by the Friedman equation in terms of ambient energy density, which is a known function of $T$, as is the temperature dependence of entropic degrees of freedom $g^s_\ast(T)$ we described above. We note that in our approach this relation is a smooth function even when the number of degrees of freedom changes as we allow for finite mass of particles introducing functions $B$ and $F$ above.

%%%%%%%%%%%%%%%%%%%%%%%%%%%%%%%%%%%%%%%%%%%%%%%%%%%%%%%%%%%%%%%%%%
\section{Chemical (Abundance) Higgs Nonequilibrium}\label{ssec:Anoneq}
In this section we discuss the magnitude of the stationary chemical nonequilibrium of the Higgs in the primordial Universe. We further present the required reaction rates allowing to obtain the non-stationary effects in the Higgs abundance in consideration of the expansion of the Universe. The Higgs particle is relatively stable, total decay width of the Higgs is $\Gamma_\mathrm{Higgs}=3.7^{+1.9}_{-1.4}$\,MeV, thus $\Gamma/m_\mathrm{Higgs}\simeq 3300$. Higgs relative stabilty is a consequence of the nature of minimal coupling which rapidly diminishes for low pair mass Higgs decay channels. The Higgs lifespan is about 500-1000 times longer compared to the top quark $\Gamma_t=1.42^{+0.19}_{-0.15}$\,GeV or gauge bosons $\Gamma_W\simeq2.0$\,GeV, $\Gamma_Z\simeq2.5$\,GeV~\cite{ParticleDataGroup:2024cfk}. 
Higgs mass ($m_h=124$\,GeV) is below the threshold that allows decay into a pair of $W^\pm$ ($ m=80.4$\,GeV) or similarly $Z^0$ ($m_Z=91.2$\,GeV). One of both of these day particles needs to be far off the mass-shell, $h\rightarrow W+W^*$, and $h\rightarrow Z+Z^*$ where $W^\ast,Z^\ast$ represent the virtual bosons~\cite{Glover:1988fn}. The branching ratios~\cite{ParticleDataGroup:2024cfk} are $53\pm 8\%$ for the bottom decay channel, $25.7\pm 2.5\%$ for the $ W W^*$-decay channel. Other noticeable decays include two gluons, $\tau$-lepton pair, $c$-quark pair, and $2.8\pm 0.3\%$ into the $ZZ^*$ decay channel. 

Higgs abundance can disappear as shown in (b)~\rf{HiggsDiagram_fig} via the decay channel $W,Z$ involving virtual particles with a branching ratio $B_{W,Z}\simeq0.285$. The virtual bosons $W^\ast$ and $Z^\ast$ decay into other particles with a coherently superposed amplitudes adding up to unity and not being hindered by on-mass individual weak interaction rates. However, the inverse reactions are incoherent, and a on-mass shell fusion reaction $WW, ZZ\rightarrow H$ is kinematically forbidden. Consequently, the inverse of the multi-particle decay (b) contributes little to the required rate to maintain equilibrium Higgs abundance. The population equation that describes the rate of change in the number of Higgs particles per unit volume is given by
\begin{align}
\label{Higgs_eq}
\frac{1}{V}\frac{dN_h}{dt}
=\sum_{i=b,c,g,\tau}(\Upsilon_i-\Upsilon_h)R_{i\overline{i}\to h}-\Upsilon_h R_{h\rightarrow W,Z},
\end{align}
where $\Upsilon_h$ is the Higgs fugacity parameter and $\Upsilon_i$ is the fugacity of the particle species $i$ and $R$ is process rate. It is convenient to define the fusion rate for the process $1+2\to3$ as follows
\begin{equation}
\Gamma(12\rightarrow 3)\equiv\frac{R(12\rightarrow 3)}{n^{th}_{3}}\,.
\end{equation}
Assuming that the back reaction for the virtual decay is insignificant the fugacity of the Higgs evolves according to 
 \begin{align}
\frac{d\Upsilon_h}{dt}\!\!
=(1-\Upsilon_h)\Gamma_\mathrm{fusion}-\Upsilon_h \Gamma_{h\rightarrow W,Z},
 \end{align}
 where the total Higgs fusion rate and decay rate are given by:
\begin{align}\label{eq:ratesDec}
\Gamma_\mathrm{fusion}=\frac{R_{b\bar{b}\rightarrow h}+R_{c\bar{c}\rightarrow h}+R_{\tau\bar{\tau}\rightarrow h}+R_{gg\rightarrow h}}{n^{th}_h},\qquad\Gamma_{h\rightarrow W,Z}=\frac{R_{h\rightarrow W,Z}}{n_h^{th}}.
\end{align}
At a given temperature the adiabatic stationary condition $d\Upsilon_h/dt=0$ allows us to obtain
\begin{align}\label{eq:chenNE}
\Upsilon_h=\frac{\Gamma_\mathrm{fusion}}{\Gamma_\mathrm{fusion}+\Gamma_{h\to W,Z}}=\frac{\Gamma_\mathrm{fusion}}{\Gamma_\mathrm{decay}}=0.69.
\end{align}
Higgs exhibits chemical nonequilibrium behavior for all $T$ in the primordial QGP.

When using the adiabatic condition required to obtain the magnitude of stationary chemical nonequilibrium~\req{eq:chenNE} we tacitly assumed that the Hubble expansion is slow compared to microscopic reaction rates. We now evaluate the rates $R$ seen in~\req{eq:ratesDec}. The invariant reaction rate per time and volume for the inverse decay reaction $1+2\to3$ has been thoroughly studied~\cite{Kuznetsova:2010pi,Kuznetsova:2008jt} 
\begin{align}
R_{12\to 3}=\frac{g_3}{(2\pi)^2}\,\frac{m_3}{\tau^0_3}\,\int^\infty_0\frac{p^2_3dp_3}{E_3}\frac{e^{E_3/T}}{e^{E_3/T}\pm1}\Phi(p_3)\,.
\end{align}
Here $\tau_3^0$ is the vacuum lifespan of particle 3. Boltzmann approximation is suitable since $m_h\ge T$, and the nonrelativistic limit when $m_3\gg T$ allows to write
\begin{align}
\Phi(p_3\to 0)=2\frac{1}{(e^{E_1/T}\pm1)(e^{E_2/T}\pm1)}.
\end{align}
The thermal decay rate per unit volume and time for heavy $m_3\gg T$ particle becomes
\begin{equation}
R_{12\to 3}=\frac{g_3}{2\pi^2}\left(\frac{T^3}{\tau_3^0}\right)\left(\frac{m_3}{T}\right)^2K_1(m_3/T)\;.
\end{equation}
$K_1$ is a Bessel function. In~\rf{Scattering_fig}, the nearly horizontal (blue) line is the total Higgs production rate from fermion fusion processes $b\bar b\to h, \tau\bar \tau\to h, c\bar c\to h$ combined with the gluon fusion process $gg\to h$. We also see the Hubble parameter (black line, multiplied by $10^{10}$) as functions of temperature. The slow expansion of the Universe allows to consider in first step the adiabatic stationary state. 

%FIGURE%%%%%%%%%%%%%%%%%%%%%%%%%%%%%%%%%%%%%%%%%%%%%%
\begin{figure}%[ht]
%\begin{center}
\centerline{\includegraphics[width=0.95\columnwidth]{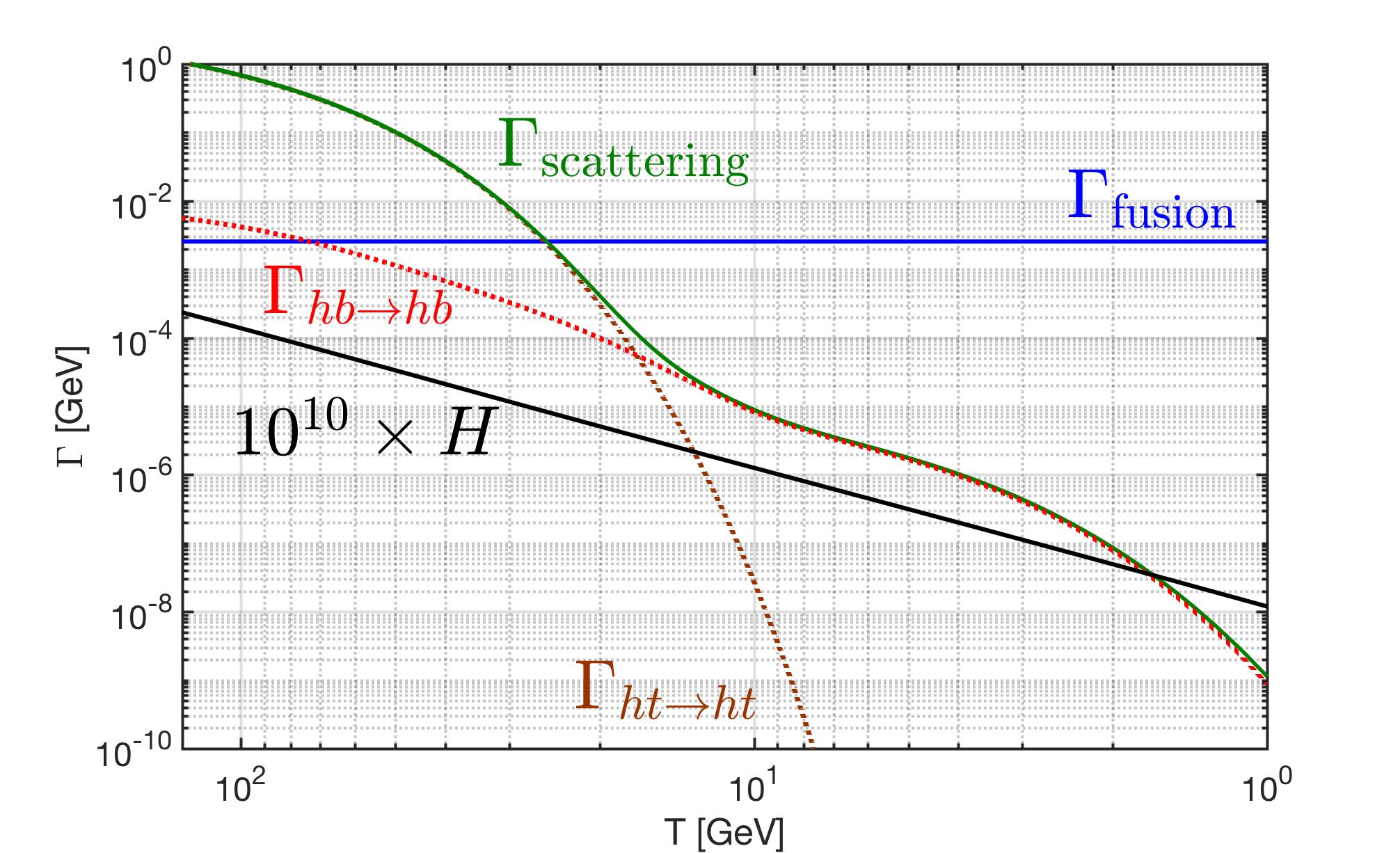}}
\caption{The total fusion rate for Higgs production (horizontal blue line) compared to the dominant scattering rates and the Hubble expansion rate (black line), scaled up with factor $10^{10}$. }
\label{Scattering_fig}
%\end{center}
\end{figure}
%%%%%%%%%%%%%%%%%%%%%%%%%%%%%%%%%%%%%%%%%%%%%%%%%%%%%

The full non-adiabatic dynamical equation for Higgs evolution requires understanding of the local density of particles and is best explored in terms of time dependence of the fugacity parameter. We consider the Higgs population equation 
\begin{align}\label{Higgs_eq002}
\frac{1}{V}\frac{dN_h}{dt}
=(1-\Upsilon_h)R_\mathrm{fusion}-\Upsilon_h R_{h\rightarrow W,Z},\qquad R_\mathrm{fusion}=\sum_{i=b,c,g,\tau}R_{i\overline{i}\to h}.
\end{align}
The left hand side in~\req{Higgs_eq002} is recast, allowing for Hubble expansion 
\begin{align}\label{number_dilution}
\frac{1}{V}\frac{dN_h}{dt}=\frac{1}{V}\frac{d(n_hV)}{dt}=\frac{dn_h}{d\Upsilon_h}\frac{d\Upsilon_h}{dt}+\frac{dn_h}{dT}\frac{dT}{dt}+3Hn_h,\;
\end{align}
where we use $d\ln(V)/dt=3H$. Substituting~\req{number_dilution} into~\req{Higgs_eq002} and dividing both sides of equation by $dn_h/{d\Upsilon_h}\simeq n^{th}_h$ (exact in Boltzmann approximation), the fugacity equation becomes
\begin{align}\label{Fugacity_eq}
\frac{d\Upsilon_h}{dt}+\Upsilon_h &\left(\frac{dn_h^{th}/dT}{n_h^{th}}\frac{dT}{dt}+3H\right)=(1-\Upsilon_h)\frac{R_\mathrm{fusion}}{n_h^{th}} - \Upsilon_h\frac{R_{h\rightarrow W,Z}}{n_h^{th}}.
\end{align}
Considering the Boltzmann limit $m_h\ge T$ for the Higgs 
\begin{align}
\frac{dn^{th}_h/dT}{n^{th}_h}\frac{dT}{dt}&=-\frac{H}{1+\frac{T}{3g^s_\ast}\frac{d\,g^s_\ast}{dT}}\left[3+\frac{m_h}{T}\frac{K_1(m_h/T)}{K_2(m_h/T)}\right]=-\frac{H}{1+\frac{T}{3g^s_\ast}\frac{d\,g^s_\ast}{dT}}\left[3+\frac{m_h}{T}\left(1-\frac{3}{2}\frac{T}{m_h}+\cdots\right)\right]\notag\\
&\approx-\frac{m_h}{T}\frac{H}{1+\frac{T}{3g^s_\ast}\frac{d\,g^s_\ast}{dT}}.
\end{align}
We now can evaluate the non-stationary effects relating to the chemical non-equilibrium. This may not be as interesting as will be the effect related to kinetic momentum distribution nonequilibrium which we discuss next. 

%%%%%%%%%%%%%%%%%%%%%%%%%%%%%%%%%%%%%%%%%%%%%%%%
\section{Higgs Nonequilibrium Momentum Distribution in the Primordial Universe}\label{ssec:Snoneq}
We now turn to the study of kinetic nonequilibrium: we obtain the scattering rates of Higgs particle in QGP using processes (${c_i}$) seen in~\rf{HiggsDiagram_fig} and show these are slower compared to the Higgs production below $T=25$\,GeV. This means that for $T<25$\,GeV the distribution in momentum of the Higgs particle is a result of the production process, and not an outcome of the following momentum distribution equilibrating scattering processes. This implies that the Higgs momentum distribution is also out of equilibrium certainly  for $T<25$\,GeV; transitioning to nonequilibrium near $T=30$\,GeV as several Higgs scattering events on top quarks are needed to achieve equilibrium momentum distribution.

In the primordial QGP, the primary interaction between quarks and the Higgs is the Compton-like scattering, see part (c) in~\rf{HiggsDiagram_fig}. The tree-level amplitudes for the threes different channels are 
\begin{align}\nonumber
 &\mathcal{M}_{c_1}=\left(\frac{m_b}{v_0}\right)^2\frac{\slashed{p}_1+\slashed{p}_2+m_b}{(p_1+p_2)^2-m^2_b}\,\,u(p_1)\overline{u}(p_4),\qquad
 \mathcal{M}_{c_2}=\left(\frac{m_b}{v_0}\right)^2\frac{\slashed{p}_1-\slashed{p}_3+m_b}{(p_1-p_3)^2-m^2_b}\,\,u(p_1)\overline{u}(p_4),\\
 &\mathcal{M}_{c_3}=\left(\frac{3m_bm_h^2}{v_0^2}\right)\frac{1}{(p_1-p_3)^2-m^2_h}\,\,u(p_2)\overline{u}(p_4),
\end{align}
where $u(p_i)$ represents the spinors for bottom quark. Using $\mathcal{M}_i$ we obtain Compton-like cross sections which lead to the angle averaged reaction rate~\cite{Letessier:2002ony}
\begin{align}
R_{12\rightarrow34}=\frac{g_1g_2}{32\pi^4}\frac{T}{1+I_{12}}\!\!\int^\infty_{s_{th}}\!\!\!\!\!\!ds\,\sigma(s)\frac{\lambda_2(s)}{\sqrt{s}}K_1(\sqrt{s}/T),\quad
\lambda_2(s)\equiv\left[s-(m_1+m_2)^2\right]\left[s-(m_1-m_2)^2\right],
\end{align}
where the cross section $\sigma(s)$ can be obtained by integrating the transition amplitude of the lowest-order Feynman diagrams, for statistical factors $g_i, I_{12}$ see Ref.\,\cite{Letessier:2002ony}. Note that all three amplitudes contribute coherently as the outcome cannot be distinguished. The Higgs-bottom/top scattering rate can be now defined as follow:
\begin{align}
 \Gamma_\mathrm{Scattering}=\frac{R_{hb\rightarrow hb}+R_{ht\rightarrow ht}}{n^{th}_h},\qquad
{\Gamma_{hq\rightarrow hq}}\equiv\frac{R_{hq\rightarrow hq}}{n^{th}_h},\qquad q=b,t.
\end{align}

In~\rf{Scattering_fig}, the rates $ht\to ht$ (dotted brown), $hb\to hb$ (dotted red) are compared with the total fusion rate for the Higgs (blue nearly horizontal solid line), and with the Hubble expansion rate (black solid line), as functions of $T$. There are several important features we can see in~\rf{Scattering_fig}: 1) In the domain of interest the Hubble rate begins to compete with scattering of Higgs below $T=2$\,GeV. However, there are practically no Higgs left in the primordial QGP so this `freeze-out' is irrelevant. 2) Below $T<17$\,GeV, we see in~\rf{Scattering_fig} that the Higgs-bottom scattering becomes the dominant kinetic process. This is so since as temperature decreases heavy particles fade out of the cosmic primordial QGP and despite their weaker coupling less massive particles dominate scattering: given the behavior of Higgs abundance we see in~\rf{HiggsDensity_fig} bottom quark is the last relevant scattering center, lighter particles contribute in nearly negligible manner at temperatures where Higgs abundance has faded out of the cosmic inventory. 

The most important feature seen in~\rf{Scattering_fig} is: 3) At the high $T$ (anti)top-quark dominated scattering rate crosses the fusion rate near to $T=25$\,GeV. Shortly before that value Higgs falls out of thermal equilibrium as it stops scattering on thermal background. Below $T=25$\,GeV we can be sure all produced Higgs cannot scatter even once before they decay. Therefore the Higgs momentum distribution remains exactly as created in the fusion process. However, as temperature decreases this kinetic shape changes due to impact of the mass thresholds. This is shown in~\rf{KineticH_fig}, where we compare as a function of $p/E$, the speed of $h$, the thermal with the kinetic production distributions: the dotted lines are the thermal distributions at $T=80,60,40,20$\,GeV., the solid lines are the corresponding kinetic production distributions which for all temperatures shown have smaller average momentum.
%FIGURE%%%%%%%%%%%%%%%%%%%%%%%%%%%%%%%%%%%%%%%%%%%%%%
\begin{figure}%[ht]
%\begin{center}
\centerline{\includegraphics[width=0.85\columnwidth]{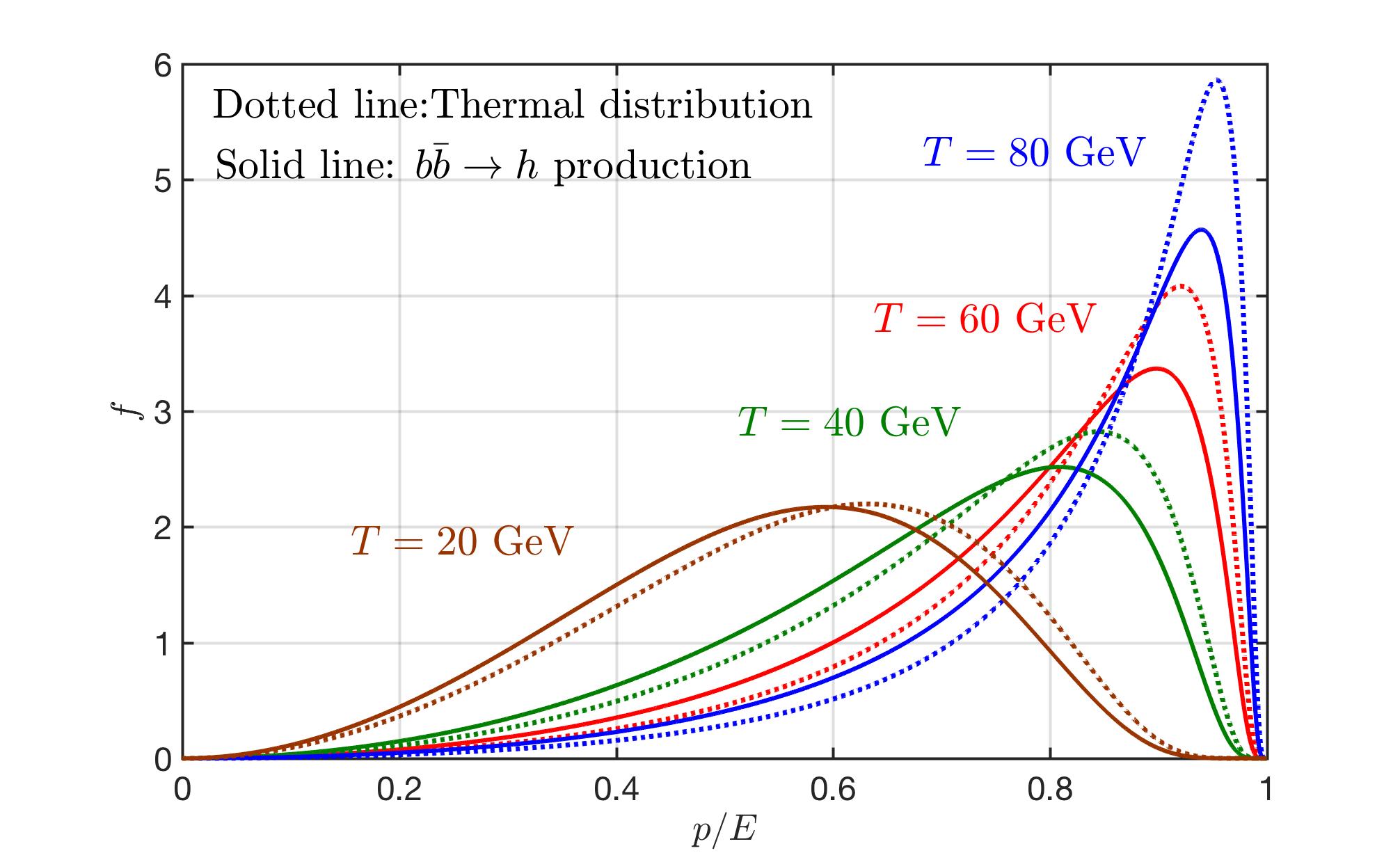}}
\caption{Higgs speed $p/E$ distributions at $T=80,60,40,20$\,GeV. The dotted line shows the thermal distribution, while the solid line represents the $b\bar b\to h$ production distribution. }
\label{KineticH_fig}
%\end{center}
\end{figure}
%%%%%%%%%%%%%%%%%%%%%%%%%%%%%%%%%%%%%%%%%%%%%%%%%%%%%

The absence of scattering equilibration with the thermal background introduces a non-stationary component into the dynamics of the Higgs during Universe evolution which we will address in the future. Of particular interest will be the domain $30>T>20$\,GeV where we expect the strongest non-stationary effects due to competition of scattering and decay rates of the Higgs. To capture this correctly aside of the use of more refined transport theory methods, such as we developed for neutrino decoupling~\cite{Birrell:2014uka}, we will need to also understand how the `melting' of Higgs vacuum structure in primordial cosmic QGP influences the decay rates we  compute here using the vacuum properties of the Higgs. To justify this effort we note in~\rf{HiggsDensity_fig} that the Higgs abundance near to $T=25$\,GeV is 6-7 orders of magnitude greater when compared to baryon asymmetry, there is ample opportunity for baryogenesis in this nonequilibrium Higgs population domain.

%%%%%%%%%%%%%%%%%%%%%%%%%%%%%%%%%%%%%%%%%%%%%%%%%%%%%%%%%%%
\section{Discussion}\label{sec:Disc}
As we have shown in this work, the Higgs particle falls out of abundance equilibrium soon after the electroweak phase transition at $T\simeq 125$\,GeV; subsequently near to $T\simeq 25$\,GeV also the Higgs momentum distribution is out of equilibrium. This means that the electroweak phase transition did not end rapidly -- indeed the duration and even the type of the phase transition could be impacted by irreversibility of the decay process into gauge mesons we described. We note that our study used vacuum particle properties but in the primordial QGP gradual melting of the electroweak vacuum especially near to electroweak phase transition could help to create a much more complex Higgs behavior we have not captured in this first study. 

This could mean that baryogenesis possible due to nonequilibrium processes at what was perceived as relatively short in time elctroweak phase transition could continue for much longer in the kinetically evolving Universe. Clearly the Higgs abundance content, and its kinetic momentum distribution, in the primordial hot Universe is a topic requiring further detailed study within full kinetic theory context allowing for dynamic vacuum properties, especially temperature dependence of $v_0=246$\,GeV, assumed at this fixed value in this work. The present work justifies this effort. This should be done not only to satisfy our curiosity: The Higgs boson is a cornerstone of the Standard Model of particle physics, and holds significant importance in understanding the fundamental forces and particles that govern our Universe. It could be a gateway to many baryon non-conserving processes. 

In this work, we examined the chemical and kinetic equilibrium of the Higgs boson during the QGP epoch in the early Universe by analyzing the relevant reaction strengths. Our findings reveal that the Higgs boson remains out of both chemical and kinetic equilibrium. The Higgs is always out of chemical abundance equilibrium with a fugacity $\Upsilon_h = 0.69$ due to decay channels into virtual gauge bosons. Additionally, Higgs momentum distribution was found to be ``cold'' for $T<25$\,GeV, since the scattering rate drops below the production rate.

Before turning to non-equilibrium processes, we have evaluated the thermal equilibrium abundance of heavy particles in QGP. Figure~\ref{HiggsDensity_fig} establishes the physical relevance of the number density ratio of heavy particles showing these as a ratio to the net baryon density in the primordial Universe, obtained under assumption of total thermal equilibrium $(\Upsilon_i = 1)$. 

In QGP, the dominant production mechanism for the Higgs boson is through the bottom-quark fusion. This process can be viewed as an inverse decay reaction ($1+2 \to 3$), where the natural decay properties of the Higgs dictate the strength of its inverse production as shown in~\rf{Scattering_fig}. In contrast, the Higgs bosons are depleted primarily by decaying into the have gauge boson pairs, which due to mass threshold includes either a virtual $W^\ast$ or $Z^\ast$. Detailed balance is broken since a decay into two Gauge bosons of which one is virtual does not have a back reaction: the inverse process is of higher order in weak interaction and is suppressed by relevant coupling constants $g^2, g^{\prime\,2}$. 

Analyzing the dominant production and decay processes of the Higgs boson in the QGP, we solve the population equation for the Higgs and demonstrate its prolonged nonequilibrium behavior, characterized by a significant departure from thermal equilibrium with $\Upsilon_h=0.69$ as a consequence of the breach of detailed balance described: This chemical nonequilibrium state arises from the dynamic balance between Higgs production via bottom-quark fusion and its decay into vector bosons $W^\ast,Z^\ast$.

There is a second type of nonequilibrium which is due to ever weaker Higgs scattering rates in the QGP: scattering on light particles can be negligible due to minimal coupling and the abundance of heavy particles decreases as temperature drops. In~\rf{Scattering_fig} we present the relevant scattering and fusion rates for the Higgs, and show that that at $T<25$\,GeV the scattering rate is smaller compared to the fusion rate. This implies that Higgs momentum distribution is governed by the production process as the particle decays before experiencing a scattering, the kinetic speed $p/E$ distributions of Higgs were presented, see~\rf{KineticH_fig}.

In conclusion, our study provides a first look at the chemical and kinetic nonequilibrium behavior of Higgs during the primordial QGP epoch. These findings not only deepen the understanding of Higgs dynamics in QGP but also offer a potential framework for future research into early Universe nonequilibrium processes allowing presence of an arrow in time required for an extended period of baryogenes.

%%%%%%%%%%%%%%%%%%%%%%%%%%%%%%%%%%%%%%%%%%%%%%%%%%%%%%%%%%%%%%%%%%
\textbf{Acknowledgment:} 
One of us (JR) acknowledges Cairns conference support by Springer-Nature, and the Southgate Foundation and the University of Adelaide, Prof. Anthony W. Thomas hospitality.

%%%%%%%%%%%%%%%%%%%%%%%%%%%%%%%%%%%%%%%%%%%%%%%%%%%%%%%%%%%%%%%%%%%
 %%%%%%%%%%%%%%%%%%%%%%%%%%%%%%%%%%%%%%%%%%%%%%

\end{document}